\documentclass{collective-intelligence}

\usepackage{graphicx}
\usepackage{array}
\usepackage{dcolumn}
\newcommand{\sigL}{\ensuremath{^{*}}}
\newcommand{\sigM}{\ensuremath{^{**}}}
\newcommand{\sigH}{\ensuremath{^{***}}}
\newtheorem{hypothesis}{Hypothesis}
\newcolumntype{d}[1]{D{.}{.}{#1}}
\newcommand{\colT}[2]{\multicolumn{#1}{c}{\textbf{#2}}}
\newcommand{\ind}[1]{\multicolumn{#1}{r}{indicators}}

\newcommand{\tablefoot}[2]{\multicolumn{#1}{l}{\small{#2}}}

\begin{document}
\bibliographystyle{agsm}

\title{COLLABORATIVE DEVELOPMENT IN WIKIPEDIA}
%
%
%
%
%

\numberofauthors{2} 
%
\author{
%
%
\alignauthor
Gerald C.\ Kane\\
       \affaddr{Boston College}\\
       \affaddr{Chestnut Hill, Massachusetts}\\
       \email{gerald.kane@bc.edu}
\alignauthor
Sam Ransbotham\\
       \affaddr{Boston College}\\
       \affaddr{Chestnut Hill, Massachusetts}\\
       \email{sam.ransbotham@bc.edu}
}
\date{3 November 2011}

\maketitle
\begin{abstract}
Using 16,068 articles in WikipediaÕs Medicine Wikiproject, we study the relationship between collaboration and quality.  We assess whether certain collaborative patterns are associated with information quality in terms of self-evaluated quality and article viewership.  We find that the number of contributors has a curvilinear relationship to information quality, more contributors improving quality but only up to a certain point. Other articles that its collaborators work on also influences the quality of an information artifact, creating an interdependent network of artifacts and contributors.  Finally, we see evidence of a recursive relationship between information quality and contributor activity, but that this recursive relationship attenuates over time.
\end{abstract}

\section{Introduction}
A new generation of IT-enabled collaborative tools --- such as wikis, blog communities, and electronic social networks --- enable people to share and create knowledge in new and potentially more powerful ways. These tools transcend traditional limitations and enable collective collaboration across conventional organizational boundaries. Nevertheless, the mere presence of these tools does not ensure effective collaboration or the creation of valuable knowledge.  People and organizations must use these tools effectively to generate valuable outcomes.

Most of the extant research on these types of IT-enabled collaboration focuses on independent peer production communities working together to produce a single information artifact such as an open source software product or a single Wikipedia article. This research overlooks the fact that these collaborative environments often produce multiple information artifacts concurrently, and contributors may transfer knowledge from one artifact to another.  Collaboration associated with the production of one information artifact may not be independent of the collaboration associated with other information artifacts on a shared collaboration platform. Understanding how  collaboration occurs on one peer-produced information artifact can be important for understanding the quality of the other artifacts produced on the shared platform. 

Wikipedia is becoming an increasingly important source of information for the general public, and it provides an excellent forum in which to examine collaborative practices \cite{KaneFichman_2009}. We examine the collaborative activities that occur between 16,068 articles and 40,479 contributors in the Wikipedia Medicine Wikiproject.  We investigate whether the quality of an article is associated with collaborative activity of its contributors to other Wikipedia articles.  Furthermore, we examine whether there is a recursive relationship between information quality and the contributions an information artifact receives.

In general, we find support for our hypotheses.  The collaborative processes that produce information artifacts on IT-enabled collaborative platforms are not independent from one another.   Instead, their development is also influenced by the work of contributors on other information artifacts on the platform.  Furthermore, we find a recursive relationship between information quality and collaboration.  More contributors create better information artifacts which in turn attracts more contributors, but this relationship attenuates over time. 

Future research examining the quality of peer produced information may be well-served by considering the interconnection between collaborative projects and the dynamics of that collaboration over time.   

\section{The Role of Information Artifacts}

Researchers have historically conceptualized IT-enabled collaborative environments as a network, examining how the structural features of those networks are associated with information benefits provided by that network \cite{Ahuja_2003,WaskoFaraj_2005}. An important difference between these previous collaborative environments and newer generations of IT-enabled collaborative tools is that people often use these emergent platforms to create peer-produced information artifacts, such as shared wiki articles, blog posts and comments, online videos and ratings, or interactive profile pages on electronic social network platforms.

These information artifacts preserve and extend the work of individual contributors and create a whole that is fundamentally different from the sum of its parts or the intentions of original contributors. A good example is how these information artifacts are often co-opted following a contributorÕs death \cite{Cohen_2009}, when they cease being an outlet for the individual and become a place to commemorate and memorialize its original author.  Thus, the peer-produced information artifacts may be used independently of the individuals who contributed to them or the purposes for which they were contributed.  As such, these information artifacts may influence and be influenced by the contributors that work on them and thus can be considered as independent entities within the collaborative platform.   

As contributors work on multiple information artifacts over time, they transfer information and knowledge contained in one artifact to another to improve the information quality of the recipient artifact. Three types of information and knowledge found in one artifact might be helpful to the development of another.  Contributors can transfer content from one artifact to another that can be used to improve the information quality of the focal artifact in a number of ways.  First, the content found in one information artifact may simply be used to improve the content of another. Second, contributors can also transfer process information from one artifact to another to improve the quality of the focal artifact.  Based on their experience working on one information artifact, contributors may have learned effective ways to collaborate with others using the IT-enabled collaborative platform.  For instance, if an individual has been involved in a large number of conflicts in other communities, s/he may have gained valuable insight on how to handle similar types of conflict in other communities. Finally, contributors can transfer reputational information about other contributors and how they contribute to (or detract from) the effective development of other information artifacts.  If one contributor has a reputation for high-quality work, other contributors may be more willing to trust their insight than another contributor who is relatively unknown.

If information artifacts have an identity independent from the contributors who create it and if the contributors can transfer information gained from one artifact to another, the result is an interconnected network of contributors and information artifacts.  The structure of this network might have significant implications for the quality of the information on the platform. Social Network Analysis (SNA) has been adopted by the organizational literature as a productive approach for studying these types of interconnected collaborative environments \cite{CrossPrusak_2002,BorgattiCross_2003,ReagansMcEvily_2003,Cummings_2004,RansbothamKaneLurie_2012}.

\subsection{Network Structure and Information Quality}

Three aspects of the collaboration that occurs in IT-enabled collaborative environments have been theorized as related to the quality of information it produces \cite{ConstantSproullKielser_1996}. 

First, the number of contributors is often associated with the quality of information. Attracting a sufficient number of contributors is important for collaborative user-generated content. More contributors increase the effort and energy dedicated to creating content and provides a broader array of knowledge and abilities for content creation. This should increase the value of collaborative user-generated content. Research on prediction markets, virtual teams, and social networks suggests that the quality of aggregate information, number of ideas generated, and likelihood of a valuable answer increases with the number of participants \cite{ConstantSproullKielser_1996,Martins_2004,FoutzJank_2010}. 

At the same time, other research suggests that having too many contributors can also be problematic. After a certain point, the marginal cost of adding new members exceeds its marginal value. Consistent with the adage ``too many cooks spoil the stew,'' an excessive number of contributors negatively influence the value of user-generated content. As the number of contributors grows, the marginal value of additional contributors decreases while the cognitive and coordination costs associated with contributions increases \cite{AsvanundClay_2004,JonesRavid_2004,RansbothamKane_2011}. In particular, those involved in the co-creation of content are likely to suffer from information overload as they try to make sense of and respond to othersÕ contributions. 

This rationale suggests that a curvilinear relationship between number of contributors and information quality. The most valuable collaborative user-generated content is generated when enough contributors are attracted to sustain production but not so much that it creates information overload for contributors. Considerable empirical evidence supports such curvilinear relationships between number of contributors and outcomes in online collaborative groups \cite{Butler_2001,HansenHaas_2001,AsvanundClay_2004,OhJean_2007,RansbothamKane_2011}. Similar relationships have also been found in traditional organizations. For instance, software development teams often need sufficient resources to accomplish their goals, but adding more members to a troubled or delayed project can compound delays by increasing coordination costs \cite{Brooks_1975}, often exponentially, as new members are added \cite{EspinosaSlaughter_2007}. Thus, we expect a curvilinear relationship between the number of contributors and the quality of collaborative user-generated content. These ideas lead to our first hypothesis:

\begin{hypothesis}\label{hyp:1}
The quality of a peer-produced information artifact will be curvilinearly (inverted-U) related to the number of contributors to the artifact.
\end{hypothesis}

Second, greater diversity of information sources provided by an IT-enabled collaborative environment improves the quality of information it produces \cite{ConstantSproullKielser_1996}.  Additional contributors may not be particularly valuable if they provide the same information already possessed by other contributors, but they are valuable when they provide access to information not already possessed by existing contributors \cite{Burt_1997,Uzzi_1997,KaneAlavi_2007}. 

The number of different information artifacts that contributors work on reflects the diversity of information available to the information artifact on a mass collaboration.  It can reflect both the knowledge directly available for transfer from other information artifacts or it may simply serve to reflect the underlying knowledge possessed by the individual contributors.  The diversity of knowledge possessed by individual contributors on an IT-enabled collaborative platform may be revealed in the pattern of other artifacts the contributors work on.  For instance, contributors with deep, specialized knowledge may work intensely in a few communities with related purposes; whereas contributors with broad, more generalized knowledge may work more superficially on a broad range of other artifacts.  

Further, these patterns reveal the type of knowledge possessed by the contributor.  For example, in the wake of the Virginia Tech Massacre, contributors reported very different reasons for contributing to the related Wikipedia article \cite{KaneFichman_2009}. Some contributors did so because of their knowledge of the school, some because they had knowledge and interest regarding the relevant gun control issues, and still others because they were skilled copyeditors.  The first type of contributor may also contribute to articles on Virginia or other colleges, the second type might also contribute to other gun-related topics, and the third type may contribute to a diverse range of articles of a particular length or stage of development.  Thus, the other information artifacts on which a contributor works on reflects the underlying knowledge and/or topical interests possessed by that contributor. We hypothesize that the number of different information artifacts on which a contributor in a peer-production community also works reflects the diversity of information sources available to the community. 
 
\begin{hypothesis}\label{hyp:2}
The quality of a peer-produced information artifact will be positively related to the number of other information artifacts on which its contributors work.
 \end{hypothesis}
 
The depth of resources available in a collaborative environment will also be related to the quality of information it produces \cite{ConstantSproullKielser_1996}.  Even if a collaborative environment provides access to a large number and to a diverse range of information sources, some sources have deeper and more valuable resources than others. Certain contributors simply provide access to greater information resources, either as a result of their connection to information artifacts with more resources or as a result of the underlying resources possessed by that individual.  Access to deeper resources generated in more active peer-production environments positively relates to the quality of information produced by the community. In many IT-enabled collaborative platforms, a relatively small percentage of information artifacts accounts for a relatively large amount of the collaborative activity that occurs on it \cite{Kuk_2006}.  Information artifacts that are the source of more collaborative activity are deeper sources of valuable content, process, and reputational information. 

Similarly, individuals who are more influential contributors to artifacts that host abundant collaborative activity may also reveal the underlying depth of resources possessed by the individual.  IT-enabled collaborative platforms typically employ limited hierarchical and administrative structures, if they possess any at all \cite{Butler_2008}.  Individuals who emerge as prominent contributors do so largely because members of the community recognize them as valuable contributors. In peer-produced information artifacts, someone can contribute heavily only by the consent of other contributors.  If other contributors do not approve of an individualÕs contributions, they will either resist them, forcing the unwelcome contributor to relent or leave \cite{Kane_2009b}; or else the other contributors will leave, as they are no longer receiving benefits from the collaborative community \cite{Butler_2001}.  

Thus, within a IT-enabled collaborative platform, the activity level of contributors on prominent information artifacts reflects the depth of resources available to a peer production community.  More active information communities are the source of deeper content, process, and reputational information than less active communities. Peer-produced artifacts with access to deeper resources are more likely to produce higher quality information.

\begin{hypothesis}\label{hyp:3}
Depth of collaborative activity that occurs within an information artifactÕs collaborative network will be positively related to artifact quality.
\end{hypothesis}

While we hypothesize above that collaboration leads to improved quality of the information artifact, it is also possible that the quality of the information artifact will also lead to certain types of collaborative behavior \cite{Kane_2009b}. As articles become of higher quality, they are more likely to attract interest from outsiders who seek to access that information, either for the content or as a collaborative exemplar. In open collaborative environments, all viewers of an article are also potential contributors.  Higher quality information may, therefore, attract more viewers, introducing the possibility that these viewers in turn become collaborators.  Thus, while we have hypothesized that certain collaborative structures lead to improved information quality, it is also possible that improved information quality will in turn lead to certain collaborative patterns.  Thus, we hypothesize a recursive relationship between the quality of an article and the collaboration it generates.

\begin{hypothesis}\label{hyp:4}
There will be a recursive, positive relationship between information quality and the collaboration that occurs on an information artifact.
\end{hypothesis}

\section{Research Method and Setting}

To test our hypotheses, we employ social network analysis (SNA). Social network analysis is capable of examining more complex networks comprising different types of nodes \cite{WassermanFaust_1994}.  A traditional but infrequently-used network conceptualization is known as the two-mode network \cite{BorgattiEverett_1997,Faust_1997}.  A two-mode network is a general network structure consisting of two fundamentally different types of entities that cannot be examined equivalently with one another. Here, we conceptualize our two-mode network as consisting of the information artifacts and individual collaborators as nodes; editing activities are the ties that connect them, due to the transfer information and knowledge from one artifact to another.  

We use two different network measures to operationalize our two remaining hypotheses --- degree centrality and eigenvector centrality.  These centrality measures are often used in conjunction to capture the features of the local (degree) and the global (eigenvector) social network \cite{Friedkin_1991,Faust_1997}.  Degree centrality in the two-mode network is used to measure the diversity of information sources available to an information artifact through its contributors. Eigenvector centrality captures the depth of information sources available to an information artifact.  This measure summarizes the node's centrality in the global network of all of the nodes and ties that compose the network.

We focus empirical analysis on the 16,068 articles within the Medicine Wikiproject in Wikipedia.  A Wikiproject is a group of contributors dedicated to develop, maintain, and organize articles related to a particular topic. We focus on a single Wikiproject, because a random sample of articles would not likely yield the social network features of theoretical interest and a Wikiproject provides clearly defined boundaries.  

\subsection{Data Collection}

We downloaded the full text history of 2,029,443 revisions of 16,068 articles by 40,479 unique contributors in the Medicine Wikiproject as of June 2009, which resulted in a 50 GB data set of raw data. We employed a 70-node Linux cluster to allow for simultaneous downloads and processing of these extensive data. For each contribution, we record the contributorÕs identity, the changes made, a description of the change, and the time of the change.

\subsection{Dependent Variables}

We use two different measures to evaluate the quality of an article.  First, we assess self-evaluated quality.  The Medicine Wikiproject evaluates articles on a 7-point scale (Stub, Start, C, B, Good, A, Featured).  We recruited two fourth year medical school students to independently validate the quality of a subsample of 120 randomly selected articles.  Each student independently evaluated each article, then ratings were compared and reconciled to create a single reviewer rating.  These reconciled ratings were then compared to the ratings assigned by the Wikiproject Medicine.  These students reached an 85\% interrater reliability with one another, and the reconciled ratings achieved a 90\% agreement with the ratings of information quality assigned by the Wikiproject.  These results suggest the self-evaluated quality was a good proxy for the overall quality of the article. 

Second, we also use public-evaluated quality in order to provide an independent and finer-grained measure of quality that provided richer data for testing Hypothesis~\ref{hyp:4}.  Here, we operationalized information quality as the number of times an article has been viewed, which is an indication of how this information is valued by the public \cite{RansbothamKaneLurie_2012}. For each article, we collected the number of views each day from December 2007 until June 2009; these data are not available for the entire history of Wikipedia. We summarized the view counts by month.

\begin{table}
\centering
\caption{Self-Evaluated Quality\label{table:1}}
\begin{tabular}{ld{4}d{4}}
\hline
\textbf{Variable}                 & \colT{1}{Model 0} & \colT{1}{Model 1} \\
\hline
Importance (medium)               &   0.532\sigH  &   0.537\sigH \\
                                  &   (0.044)   &   (0.044) \\
Importance (high)                 &   0.785\sigH  &   0.736\sigH \\
                                  &   (0.080)   &   (0.082) \\
Importance (top)                  &   1.383\sigH  &   1.251\sigH \\
                                  &   (0.152)   &   (0.161) \\
Age                               &   0.137\sigH  &   0.103\sigM \\
 \hspace{0.1in} (ln, years)       &   (0.036)   &   (0.036) \\
Daily Views                       &   0.218\sigH  &   0.180\sigH \\
 \hspace{0.1in}(ln, views)        &   (0.033)   &   (0.033) \\
Length                            &   1.876\sigH  &   1.753\sigH \\
  \hspace{0.1in} (ln, characters) &   (0.094)   &   (0.097) \\
Complexity (ARI)                  &   -4.191\sigH &   -4.366\sigH \\
                                  &   (0.488)   &   (0.493) \\
Section Depth                     &   0.144\sigH  &   0.135\sigH \\
                                  &   (0.028)   &   (0.028) \\
External References               &   0.253\sigH  &   0.221\sigH \\
                                  &   (0.032)   &   (0.030) \\
Multimedia content                &   -0.063    &   -0.067 \\
                                  &  (0.048)    &   (0.051) \\
Anonymity                         &   0.905\sigH  &   1.029\sigM \\
 \hspace{0.1in} (percentage)                    &  (0.149)    &   (0.149) \\
Distinct Contributors             &             &   0.269\sigH \\
                                  &             &   (0.044) \\
Distinct Contributors             &             &   -0.186\sigH \\
 \hspace{0.1in} (squared)         &             &   (0.030) \\
Degree Centrality                 &             &   0.168\sigH \\
   \hspace{0.1in} per Contributor               &             &   (0.034) \\
Eigenvector Centrality            &             &   0.105\sigH \\
                                  &             &   (0.022) \\
\hline
Log pseudolikelihood              &   -9,036.96  &   -8,949.33 \\
Wald $\chi^2$                     & 4,294.34\sigH  &   4,332.24\sigH \\
Pseudo $R^2$                      &   36.74     &   37.30 \\
\hline
\end{tabular}
\end{table}

\begin{table*}
\centering
\caption{Three Stage Least Squares Model of Article Views\label{table:2}}
\begin{tabular}{lld{5}d{5}d{5}d{5}} 
\textbf{Model} & \textbf{Variable}  & \colT{2}{Model 1}  & \colT{2}{Model 2} \\
\hline
\multicolumn{4}{l}{\textbf{Equation 1: Article Views (ln/1000)}}  \\
               & Monthly Fixed Effects                          &  \ind{1}       &   & \ind{1} \\ 
               & Constant                                       & 744.439\sigH   & (5.215)       & 755.126\sigH   & (5.387)   \\
               & Article Views (ln, lagged)                     &   0.971\sigH   & (0.001)       &   0.971\sigH   & (0.001)   \\
               & Age (ln, years)                                &  18.287\sigH   & (1.285)       &  17.964\sigH   & (1.384)   \\
               & Length (ln, characters)                        &   9.135\sigH   & (1.164)       &   8.904\sigH   & (1.166)   \\ 
               & Complexity (ARI)                               & -27.702\sigL   &(11.181)       & -26.060\sigL   &(11.181)   \\
               & Section Depth                                  &   1.975\sigL   & (0.847)       &   1.795\sigL   & (0.848)   \\
               & External References                            &   1.038        & (0.690)       &   0.629        & (0.693)   \\
               & Internal Links                                 &   3.261\sigM   & (1.129)       &   2.990\sigM   & (1.135 )   \\
               & Multimedia Content                             &  -0.014        & (0.661)       &  -0.031        & (0.661)   \\
               & Anonymity (percentage)                         &  91.676\sigH   & (5.857)       &  75.019\sigH   & (6.171)   \\
               & Contributors                                   &  12.280\sigH   & (1.424)       &  48.579\sigH   & (4.627)   \\
               & Contributors$^2$                               &  -6.780\sigH   & (1.115)       & -61.869\sigH   & (6.901)   \\
               & Local Centrality                               &  -1.993\sigM   & (0.672)       &  -2.738\sigH   & (0.729)   \\
               & Global Centrality                              &   3.266\sigH   & (0.685)       &   1.908\sigL   & (0.765)   \\
               & Age $\times$ Contributors                      &                &               & -27.487\sigH   & (3.483)   \\
               & Age $\times$ Contributors$^2$                  &                &               &  42.469\sigH   & (5.158)   \\
               & Age $\times$ Local Centrality                  &                &               &  -0.856        & (0.747)    \\
               & Age $\times$ Global Centrality                 &                &               &  -1.694\sigH   & (0.391)    \\
               \cline{2-6} \noalign{\smallskip}
               & $R^2$                                          &   98.099       &               &  98.100         \\
               & $\chi^2 (\times 10^6)$                         &    6.44\sigH   &               &   6.44\sigH     \\ \noalign{\smallskip}
\hline
\multicolumn{4}{l}{\textbf{Equation 2: Contributors}}  \\
               & Monthly Fixed Effects                          & \ind{1}        &               & \ind{1}        \\
               & Constant                                       &  -0.250\sigH   & (0.053)       &  -0.251\sigH  & (0.053)    \\
               & Contributors (lagged)                          &   1.021\sigH   & (0.001)       &   1.021\sigH  & (0.001)    \\
               & Article Views (ln)                             &   0.103\sigH   & (0.006)       &   0.103\sigH  & (0.006)    \\ 
               & Age (ln, years)                                &  -0.121\sigH   & (0.011)       &  -0.121\sigH  & (0.011)    \\
               & Length (ln, characters)                        &   0.002        & (0.012)       &   0.002       & (0.012)    \\
               & Complexity (ARI)                               &  -1.034\sigH   & (0.113)       &  -1.034\sigH  & (0.113)    \\
               & Section Depth                                  &   0.041\sigH   & (0.009)       &   0.041\sigH  & (0.009)    \\
               & External References                            &  -0.052\sigH   & (0.007)       &  -0.052\sigH  & (0.007)    \\
               & Internal Links                                 &   0.139\sigH   & (0.011)       &   0.139\sigH  & (0.011)    \\
               & Multimedia Content                             &  -0.025\sigM   & (0.007)       &  -0.025\sigM  & (0.007)    \\
               & Anonymity (percentage)                         &   0.407\sigH   & (0.056)       &   0.407\sigH  & (0.056)    \\
               & Article Protected? (1=yes)                     &  -5.787\sigH   & (0.087)       &  -5.789\sigH  & (0.087)    \\
               \cline{2-6} 	\noalign{\smallskip}
               & $R^2$                                          &  99.958        &               &  99.958     \\
               & $\chi^2 (\times 10^8)$                         &   3.02\sigH    &               &   3.02\sigH     \\
\hline
\tablefoot{6}{124,711 observations; standard errors in parentheses; significance $\sigL p<0.05$, $\sigM p<0.01$, $\sigH p<0.001$.}
\end{tabular}
\end{table*}

\section{Analysis and Results}

The full dataset was then analyzed using Stata.  Ordinal regression is appropriate when there is a progressive relationship within a categorical dependent variable, but it is unclear the magnitude of different between the categories.  For instance, the observer may know which Olympic athletes have won the gold, silver, and bronze medals without knowing the final scores of any of the athletes.  This method is most appropriate for our measure of self-evaluated quality. Table~\ref{table:1} describes the full results of an ordinal logistic regression on self-evaluated quality. Model 0 presents our results for the baseline model with only control variables, and Model 1 presents the results of our models with the variables of interest.  (We also tested each independent variable of interest independently, and results are consistent with the composite findings in Model 1.)  

Examining Hypothesis~\ref{hyp:1}, both the linear and square coefficients are significant ($\beta = 0.269$, $p < 0.001$ and $\beta = -0.186$, $p < 0.001$, respectively).  These coefficients indicate an inverted-U relationship with article quality.  Additional contributors working on an information artifact increase quality up to an optimal point, but then additional contributors detract from the quality of information found in the artifact. We also find support for the second hypothesis that diversity of resources increases artifact quality.  The coefficient on degree centrality per contributor is positive and significant ($\beta = 0.168$, $p < 0.001$).  The more diverse the content, process, and reputational knowledge contributors access and/or represent, the higher quality the information artifact is likely to be. Hypothesis~\ref{hyp:3} that depth of recourses available to an information artifact will be positively related to information quality is also supported.  The coefficient on eigenvector centrality is positive and significant ($\beta = 0.105$, $p < 0.001$).  The greater depth of resources that an artifact can access through or is represented by its contributors, the higher quality the artifact is likely to be.

Table~\ref{table:2} describes the results of a simultaneous equation, three stage least squares regression on the natural log of article views (scaled by 1,000 for presentation) using the sample of 124,711 monthly observations of articles from December 2007 until June 2009. Model~1 introduces the focal network variables. Because of the large sample size, we use a low threshold of statistical significance ($p < 0.001$) to test our hypotheses. 

We find partial support for Hypothesis~\ref{hyp:4}.  The number of unique contributors to user-generated content has a curvilinear relationship with article views. Both the linear and squared coefficients are significant ($\beta = 12.28$, $p < 0.001$ and $\beta = -6.78$, $p < 0.001$, respectively). These coefficients indicate an inverted-U relationship with article views. Additional contributors working on an article increase its quality up to an optimal point, but then detract from the ability of the article to attract viewers. We also considered models that utilized either a linear effect of contributors or a log of the number of contributors; based on the Akaike information criterion (AIC); the quadratic model provides a slightly better fit (decreases in AIC of 18 and 7 respectively). 

However, we do not find that the quality of user-generated content is positively related to local network centrality; the coefficient for degree centrality per contributor is negative and significant ($\beta = -1.99$, $p < 0.010$).  We do find that the quality of user-generated content is positively related to global network centrality; the coefficient for eigenvector centrality is positive and significant ($\beta = 3.27$, $p < 0.001$). As hypothesized, both models demonstrate a recursive effect of article viewership on the number of contributors. In Equation~2, the coefficient for article views is significant and positive ($\beta = 0.10$, $p < 0.001$). More contributors lead to greater viewing, but more viewing also yields a greater number of contributors. The protect variable, used for identification of the simultaneous model since it affects contributions but not viewing, is also significant ($\beta = -5.79$, $p < 0.001$) and behaves as expected. When an article has restrictions placed on who can contribute, significantly fewer people contribute to it. 

It is interesting to note that age has the opposite effect in the contributor model --- age is positively related to viewing but negatively related to the overall number of contributors. This suggests that collaborative user-generated content matures and stabilizes over time; more people come to view older content but they are less likely to contribute to that content. It may be that more mature content attracts a more general audience that is less likely to have the knowledge or inclination to contribute, or it may be that the viewers of the content find it to be relatively complete and feel they have nothing to add to improve it.

\section{Conclusion}

In this paper, we test the influence of the network structure created by contributors and information artifacts on information quality in peer-produced information.  We find good general support for our hypotheses, determining that the collaboration occurring on one information artifact can influence the quality of the other information artifacts on which those collaborators work.  Furthermore, we also demonstrate a recursive relationship between contributors and information quality in terms of viewership.  More viewers brings more collaborators which brings more collaborators, but this recursive relationship attenuates over time.  Implications are that researchers should broaden their understanding of how collaboration on other information artifacts can influence information quality and begin understanding peer-production settings as a network of knowledge.

\section{Acknowledgments}
Gerald Kane acknowledges funding for this research from the National Science Foundation (CAREER 0953285).

\bibliography{General}  
%

\end{document}